\documentclass{article}
\usepackage{spconf}
\usepackage{amsmath,graphicx}
\usepackage{booktabs}
\usepackage{comment}
\usepackage{cite}
\usepackage{multirow,multicol}
\usepackage{url}

\title{Improving sound event detection metrics: insights from DCASE 2020}
\name{
\begin{tabular}{@{}c@{}}
    Giacomo~Ferroni, Juan~Azcarreta,\\Francesco~Tuveri, \c{C}a\u{g}da\c{s}~Bilen\\Sacha~Krstulovi\'{c} \\ \\
    {\em Audio Analytic - AA Labs},\\{\em 2 Quayside,}\\{\em Cambridge, CB5 8AB, UK}\relax
\end{tabular}\hskip 1in
\begin{tabular}{@{}c@{}}
    Nicolas~Turpault,\\Romain Serizel\\\thanks{UL/INRIA's work for this article was partly supported by the French National Research Agency (project LEAUDS “Learning to under-stand audio scenes” ANR-18-CE23-0020) and by the French region Grand-Est.} \\ \\
    {\em Universite de Lorraine, CNRS,}\\{\em Inria, Loria,}\\{\em F-54000 Nancy, France}\relax
\end{tabular}
}
\address{}
\begin{document}
\maketitle
\begin{abstract}
The ranking of sound event detection (SED) systems may be biased by assumptions inherent to evaluation criteria and to the choice of an operating point. This paper compares conventional event-based and segment-based criteria against the Polyphonic Sound Detection Score (PSDS)'s intersection-based criterion, over a selection of systems from DCASE 2020 Challenge Task 4. It shows that, by relying on collars, the conventional event-based criterion introduces different strictness levels depending on the length of the sound events, and that the segment-based criterion may lack precision and be application dependent. Alternatively, PSDS's intersection-based criterion overcomes the dependency of the evaluation on sound event duration and provides robustness to labelling subjectivity, by allowing valid detections of interrupted events. Furthermore, PSDS  enhances the comparison of SED systems by measuring sound event modelling performance independently from the systems' operating points.
\end{abstract}
\section{Introduction}
\label{sec:intro}

Given an audio  recording, sound event detection consists not only of detecting which sound events have occurred during the recording but also when it occurred~\cite{virtanen2018computational}. SED and the broader domain of ambient sounds analysis have received a lot of attention in the past few years because of their numerous potential applications including smart home consumer electronics, smartphones, earbuds, assisted living, smart cities or security~\cite{Bello:SONYC:CACM:18,radhakrishnan2005audio,serizel2016, zigel2009method, AA:hearables2019, 6529487, AA:smarthome2018}.

A common way to solve the SED challenge is to use supervised machine learning on a training dataset composed of strongly labeled recordings, i.e., with timestamps for the sound events' onset and offset. However, obtaining these annotations is time and resource consuming, as well as prone to errors and disagreement between annotators in their perception of where the onsets and offsets should be, thus making strong annotations difficult to obtain from crowd sourcing. Since 2018, Task 4 of the international challenge on Detection and Classification of Acoustic Scenes and Events (DCASE) has been addressing this issue by using weakly annotated training data~\cite{serizel2018_DCASE} and synthetic soundscapes with strong annotations~\cite{Turpault2019_DCASE}, which are much cheaper to obtain. However, the evaluation set remains strongly annotated by humans, with potential inconsistencies in annotations due to the aforementioned disagreements and ambiguities, whereas the systems are evaluated on a segmentation task, i.e., in terms of their capacity to localise a sound event in time in addition to predicting the correct sound event classes. Therefore, evaluation metrics should be able to measure the time-localisation abilities of the compared systems at different time scales while remaining independent of the labeling strategy.
With that in mind, and until recently, SED systems were evaluated on two different classes of metrics summarised into F1-scores~\cite{giannoulis:hal-01123764,Mesaros:2016}: event-based metrics, which measure how successful the models are in correctly predicting each annotated sound event, or segment-based metrics, which measure how successful the models are in correctly identifying the most dominant sound event within fixed sized segments of audio.

In 2020, DCASE Challenge Task 4 introduced the Polyphonic Sound Detection Score (PSDS)~\cite{PSDS:2020} as an alternative evaluation metric~\cite{turpault:hal-02891665,turpault:hal-02891700}. PSDS relies on an intersection criterion to validate the predicted sound events against the reference annotations. It can be adjusted to enforce similar level of constraints to segment or event based evaluations and it can provide the advantages of both approaches (Table~\ref{tab:diff_summary}). PSDS also provides an alternative metric to the F1-score insofar as it is a performance figure that is global across a set of operating points, rather than depending on a single operating point chosen a priori for each submitted system as implied by the F1-score. Therefore PSDS prevents the erroneous conclusion that a system performs sound event modelling better than another one where, instead, the difference relies only on a different choice of system sensitivity tuning.

This paper therefore presents the experimental comparison, over PSDS compliant submissions to DCASE 2020 Challenge Task 4,
of the intersection-based metric versus the conventional event or segment-based metrics, with particular regards to flexibility, robustness to annotation variations and independence from the operating point.

Section~\ref{sec:background} describes the compared SED evaluation metrics and the analysis methodology used in this paper. Section~\ref{sec:criteria} and Section~\ref{sec:operating_point} report the experimental results which outline the relevance of the choice of evaluation criterion and the independence of the PSDS score from the operating point, respectively. Finally, the conclusions of the paper are summarised in Section~\ref{sec:conclusions}.


\section{The Compared SED Evaluation Criteria}
\label{sec:background}

Several metrics have been proposed for the evaluation of polyphonic SED systems~\cite{Mesaros:2016}. This section presents the metrics that have been used in DCASE Challenge Task~4 during the past 4 years as well as the metric recently introduced and presented in~\cite{PSDS:2020}.
In particular, this paper focuses on two aspects of importance for the evaluation of SED systems: different possible ways to validate the predicted time localisation of a sound event, either at a segment level or at an event level, and different approaches available to compute an overall performance metric from the validated event-wise or segment-wise decisions.
%
%
%
%
%

\begin{table}[ht]
    \centering
    \begin{tabular}{p{5.5cm}ccc}    \toprule
    & \emph{SB} & \emph{CB} & \emph{IB}  \\\midrule
    Allows interrupted detections & yes & no & yes \\
    Allows merged detections & yes & no & yes \\
    Each event has an equal  & \multirow{2}{*}{no} & \multirow{2}{*}{yes} & \multirow{2}{*}{yes} \\
    weight in the metric & & &\\
    Metric parameters can be chosen  & \multirow{2}{*}{no} & \multirow{2}{*}{no} & \multirow{2}{*}{yes}\\
    independently of event durations & & &\\
     \bottomrule
     \hline
    \end{tabular}
    \caption{Compared characteristics of the segment-based (SB), collar-based (CB) and intersection-based (IB) approaches.}
    \label{tab:diff_summary}
\end{table}

\subsection{Event validation criteria}
Time localisation can be determined either by comparing each predicted sound event with the reference annotations in the case of event-based metrics, or by determining the presence of a sound event within a segment of fixed length in the case of segmented-based metrics.
Although intersections and collars are both event-based criteria, in this article the two approaches are distinguished as \emph{intersection-based} and \emph{collar-based} for the sake of comparison.

\textbf{Collar-based criterion}:
Traditional event-based metrics match predicted event occurrences with reference events in the transcription up to a tolerance collar around the onset and the offset. These collars generally have a fixed size for the onset, e.g., 200ms in DCASE Challenge Task~4. The size of the collar on the offset can depend on the duration of the sound event, e.g., the maximum between 200ms and 20\% of the duration of the sound event in DCASE Challenge Task~4. As a consequence, onset detection can have a large impact on event-based metrics. For short sound events, typically below 1 second in length, it could seem fair to impose a constraint of 200ms on onset estimation. However, for longer sound events the impact of this collar might become disproportionate.

\textbf{Segment-based criterion}:
Segment-based metrics are computed by comparing the system outputs against the reference across segments of fixed duration.
This approach is more tolerant than event-based metrics to the occurrence of short pauses between consecutive sound events as well as to errors in the detection of sound event boundaries~\cite{serizel:hal-02114652}.
The segment-based criterion requires to define a fixed segment size, e.g., 1 second for DCASE Challenge Task~4. This value is application-dependent and has to be chosen carefully in order to avoid performance measure artifacts. Indeed, multiple short sound events (dog barks, dishes, etc.) present in the same segment may be artificially collapsed into a single target event, thus leading to an overestimation of performance. Conversely, long sound events (vacuum cleaner, blender, etc.) may be artificially subdivided into sequences of sound events, one per segment, thus possibly underestimating performance and overemphasising the total contribution of long event detections to the final performance figure.

\textbf{Intersection-based criterion}:
The intersection criterion is a novel approach, formally presented in \cite{PSDS:2020}. It redefines true positives (TP), false positives (FP) and cross-triggers (CT) in the context of polyphonic SED
by exploiting the amount of overlap between the annotations and the detections. Thus, it allows the evaluation to support merged detections, i.e., single detections across multiple annotations, and interrupted ones, i.e., multiple detections across single annotations. Furthermore, the evaluation is invariant to the time scale of the event durations. These properties provide the flexibility needed to handle subjective labelling and to avoid event duration biases, i.e., events with certain durations have a higher weight in the evaluation.
The minimum amount of overlap required to validate the detections against the annotations is parameterised by three intersection ratios: the Detection Tolerance Criterion (DTC), the Ground Truth intersection Criterion (GTC) and the Cross-Trigger Tolerance Criterion (CTTC)~\cite{PSDS:2020}. 

A summary of the main differences between the three event validation criteria is shown in Table~\ref{tab:diff_summary}.

\subsection{Overall performance metrics}
Once a criterion has been chosen to determine the validity of the time localisation for each predicted sound event, a metric needs to be computed from the predictions of the system at the corresponding time scale.

\textbf{F1-score}:
SED systems are conventionally evaluated with an event error rate (ER) or an F1-score~\cite{Mesaros:2016}. The computation of these metrics relies on the evaluation of TP and FP rates computed from system predictions made at a single operating point. The issue about computing metrics on a single operating point will be addressed in Section~\ref{sec:operating_point}.

\textbf{ROC-curves/PSDS}:
Receiver Operating Characteristic (ROC) curves are computed over ranges of operating points. As such, the Area Under ROC Curves (AUC)~\cite{Tatbul:2018} summarises global performance information without depending on the choice of a single operation point. However, the AUC is traditionally used for binary classification. PSDS defines a similar metric, but for polyphonic SED computed from a ROC-like curve~\cite{PSDS:2020}.

\subsection{Analysis methodology}
\label{sec:methodology}
In this analysis, we used the PSDS compliant systems submitted to the DCASE 2020 Challenge Task 4. In total, eight teams were selected for the analysis\footnote{The evaluation data presented in this paper have been computed by the Task 4 organisers (UL/INRIA) with code provided by Audio Analytic, in order to preserve the confidentiality of the challenge's prediction data.}, and only the best submission for each team in terms of PSDS score computed with $\alpha_{CT}=\alpha_{ST}=0$, DTC=GTC=0.5, CTTC=0.3 and $e_{max}=100$ events-per-hour~\cite{PSDS:2020}. The dataset used throughout the comparison is the domestic environment sound event detection (DESED) public evaluation dataset~\cite{Turpault2019_DCASE, Serizel2020_ICASSP}. It is entirely composed of strongly annotated, real soundscapes and it comprises the 10 different sound event classes shown in Table~\ref{tab:publicevaldata}.

The predictions from the selected systems are evaluated using the metrics presented in Section~\ref{sec:background} with 200ms and $\max(200\text{ms}, 0.2*\text{annotation\_length})$ as onset and offset collars, respectively. The segment size for segment-based evaluation is set to 1 second~\cite{Heittola:2013}. The intersection-based evaluation uses DTC=GTC values ranging from 0.1 to 0.9 with a step of 0.2 while the CTTC threshold is not considered since $\alpha_{CT}$ is set to 0.
The analysis presented in the next two sections aims to study how the different evaluation criteria compare to each other, and how evaluating systems independently from the operating point keeps the evaluation focused on sound event modelling.

\begin{table}[ht]
    \centering
    \begin{tabular}{p{1.25cm}c|p{1.65cm}c}    \toprule
    \multicolumn{2}{c}{Short events} & \multicolumn{2}{c}{Long events} \\
    \emph{Class} & \emph{$\mu\pm\sigma$, $m$} & \emph{Class} & \emph{$\mu\pm\sigma$, $m$}  \\\midrule
     Dishes & 0.6±0.7, 0.33 & El. shaver & 5.4±2.1, 5.76\\
     Dog & 0.7±3.3, 0.39 & Blender &	5.6±2.9, 5.96\\
     Cat & 1.1±1.0, 0.74 & Run. water & 6.4±1.1, 7.85\\
     Speech & 1.4±2.2, 1.04 & Frying & 8.7±3.6, 10.00\\
     Alarm & 2.4±2.9, 0.83 & Vacuum & 8.3±0.6, 9.96\\
     \bottomrule
     \hline
    \end{tabular}
    \caption{Class names and label duration statistics: mean ± standard deviation and median in seconds.}
    \label{tab:publicevaldata}
\end{table}

\section{Comparing event validation criteria}
\label{sec:criteria}
Throughout this section the metric is set to macro F1-score, computed as the average of the per-class F1-scores, and is used to compare the different event validation criteria.

\begin{figure}[t]
  \centering
  \centerline{\includegraphics[width=\linewidth]{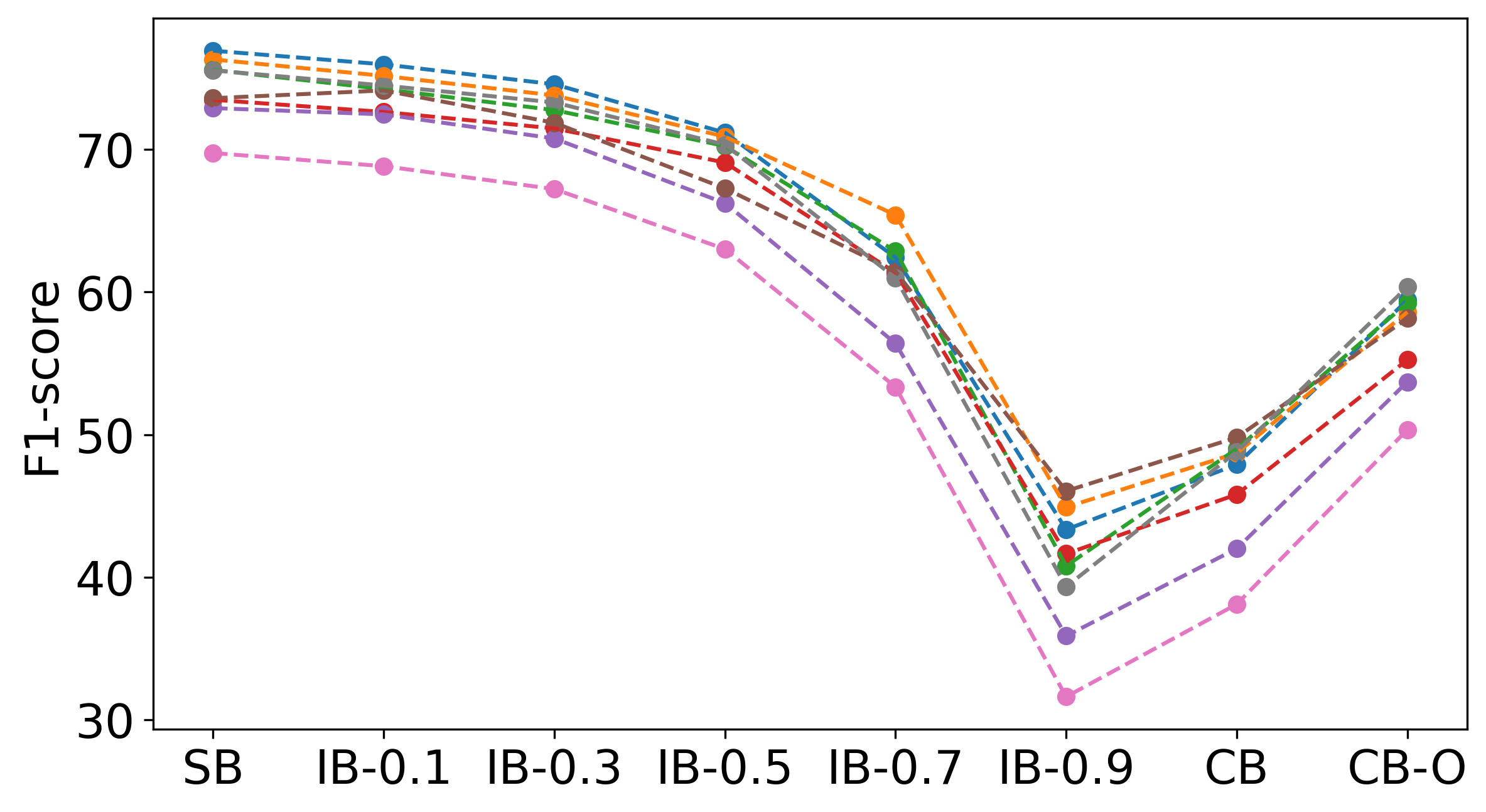}}%
    \vspace{-0.5cm}
\caption{Macro F1-score across systems for the compared evaluation criteria: segment-based (SB), intersection-based (IB-DTC=GTC), collar-based (CB) and collar onset only (CB-O).}
\label{fig:F_COMPARISON}
\vspace{-4mm}
\end{figure}

From Fig.~\ref{fig:F_COMPARISON}, when the intersection criterion is defined by GTC=DTC=0.1 (IB-0.1) it produces similar F1-scores to the segment-based (SB) criterion, while GTC=DTC=0.9 (IB-0.9) behaves similarly to the onset/offset collar-based (CB) criterion. Thus, by changing the GTC/DTC parameters, it is possible to move smoothly between these two extreme constraints. This leads to an absolute variation of about 20\% to 30\% in the value of the reported F1-score.

\begin{figure*}
  \centering
  \centerline{\includegraphics[width=\linewidth]{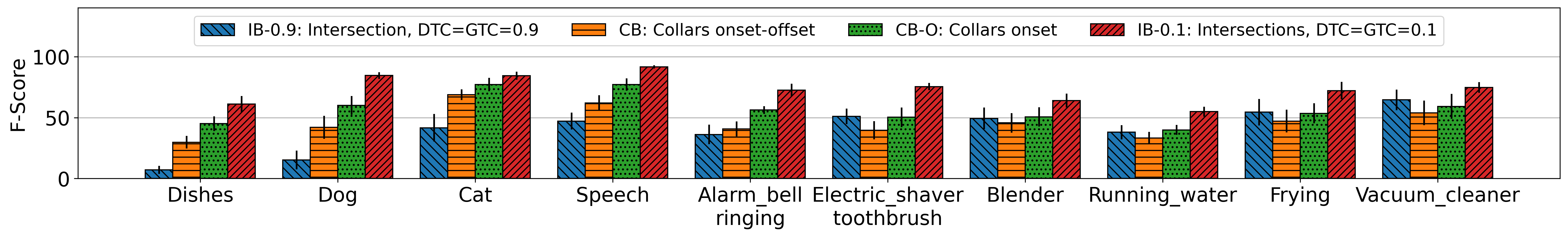}}%
  \vspace{-0.5cm}
\caption{Per-class F1-score comparison of event-based metrics, with standard deviations, across the eight systems.
}
\label{fig:BARCHART}
\vspace{-4mm}
\end{figure*}

For applications where the end-user only needs to be notified about the occurrence of a sound instance, the evaluation constraints can be relaxed when it comes to detecting the exact time boundaries of the event.
Indeed, when setting the DTC/GTC parameters to a low value, the evaluation relies on the existence of a detected sound event that has some overlap with the ground truth, rather than on the precise match of event localisation or event duration, to decide if a detection should be considered as a true positive.
In this case, a system which produces short detections but is always correct is a very good system.
Conversely, setting DTC/GTC to high values rewards the systems which almost perfectly match the ground truth boundaries, but comes back to suffering from labeller's subjectivity and lack of clear definitions for the location of sound event boundaries.
An evaluation criterion that is robust to these limitations is therefore preferable, and setting high DTC/GTC values is not recommended.
A value of DTC=GTC=0.5 seems suitable because it represents a trade-off between enforcing accurate localisation and handling annotation subjectivity, including interrupted detections and single detections over multiple ground truths, a trade-off that the conventional evaluation criteria are not able to achieve.

Furthermore, collars introduce a dependency on the annotation lengths: event durations comparable to the collar size imply more tolerance on onset detection, e.g., in the case of 500ms-long events with a 200ms collar. Comparatively, the same tolerance applied to 8-seconds-long events is stricter because it ignores a lot of the matching information, in particular the possibility for valid detections to occur outside of the collar boundaries but within the whole event duration.

Three observations confirm this from Figure~\ref{fig:BARCHART}, which shows the F1-scores per-class averaged across the eight systems for the compared evaluation criteria, in growing order of class duration from Table~\ref{tab:publicevaldata}:
(1) IB-0.9 always yields lower F1-scores than at IB-0.1 (blue vs red), i.e., the strictness of the validation of detections is proportional to the required amount of overlap.
However, IB-0.9 always yields higher F1-scores on long event classes than CB, and vice versa for short event classes,  (blue vs orange) thus confirming that the collars are imposing a stricter validity constraint, perhaps unfairly, on long classes than on short classes.
(2) CB always yields lower F1-scores than CB-O (orange vs green). This is expected, since the former case is a more relaxed constraint because offset detection is not required. However, the performance delta is lower for long events than for short events: offset validation has less influence on the perceived performance for long events than for short ones.
(3) IB-0.9 versus CB-O (blue vs green) yields very similar F1-scores for long events only, meaning that validating the onset for long events is a difficult task, i.e., as difficult as generating detections that almost fully overlap with the corresponding annotations.

\section{Independence from operating point}
\label{sec:operating_point}

\begin{figure}[t]
  \centering
  \centerline{\includegraphics[width=\linewidth]{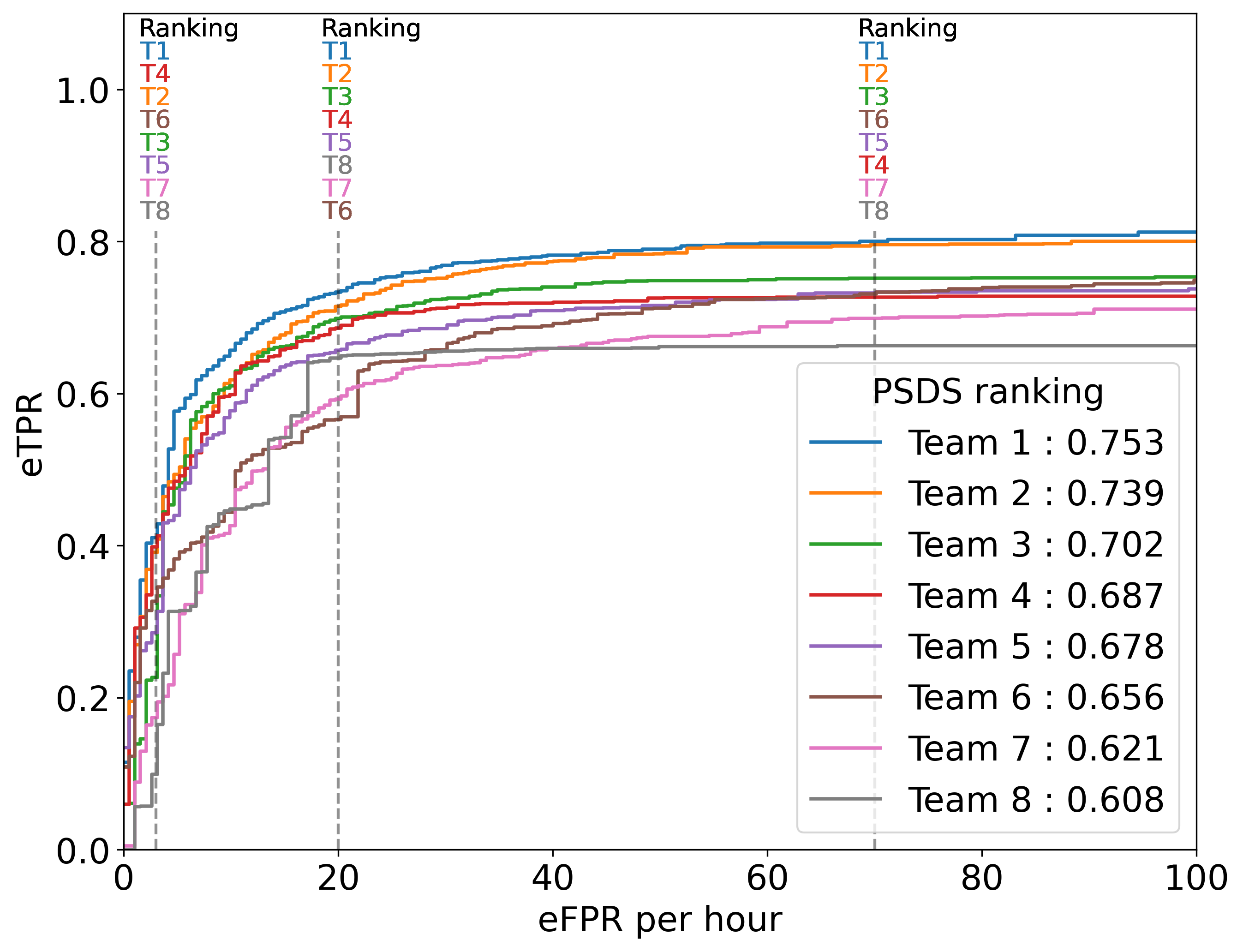}}%
    \vspace{-0.5cm}
\caption{PSD-ROC curves per team. The team ranking varies depending on the operating point. Comparing systems across whole PSD-ROC curves factors out this dependency.}
\label{fig:PSD_ROC_rankings}
\vspace{-4mm}
\end{figure}

The F1-score evaluation requires participants to tune thresholds that optimise the compared systems for a specific cost of false positives and missed detections~\cite{Krstulovic:2018}. This process can have a considerable impact on system ranking, as illustrated in Figure~\ref{fig:PSD_ROC_rankings}. For example, when selecting a system for a low FP use-case (leftmost ranking), system T4 would be preferred to system T3. However, in a scenario where FPs have a lower cost (rightmost ranking), T3 would be a better candidate model than T4.
Scientific evaluation tasks such as DCASE 2020 Challenge Task~4 are interested in finding the best modelling method in general, regardless of user-related specifications of error costs and trade-offs.

From that standpoint, the area under the PSD-ROC, which averages performance across all possible operating points, delivers a single evaluation metric that is independent of use-case scenarios and allows the ranking to be representative of general sound event modelling performance rather than operating point optimisation.

\section{Conclusions}
\label{sec:conclusions}
This paper analyses the influence of evaluation metrics in comparing the performance of SED systems in the context of DCASE 2020 Challenge Task~4, and considers two factors leading to comparison bias: the event validation criterion and the dependency on the operating point.
Sound event validation based on the collar-based criterion treats events of different duration with different levels of strictness. Alternatively, the segment-based criterion appears more tolerant to localisation errors but it lacks precision and may still be application dependent because the chosen time resolution may overemphasise the contribution of particular event lengths to the final performance figure, while de-emphasising others. A solution is shown to be the intersection-based criterion, which adapts to the audio event duration by design, does not focus on annotation boundaries nor on defining a fixed time resolution for the evaluation, and provides robustness to labelling subjectivity by allowing valid detections across interruptions.
Once an event validation criterion is chosen, system rankings based on the F1-score are still locally biased by multiple choices of operating point: PSDS overcomes this dependency. Overall, this paper suggests that PSDS, powered by the intersection-based criterion, captures a more direct relationship between performance improvements and sound event modeling methods than the conventional evaluation metrics.

\bibliographystyle{IEEEbib}
\bibliography{refs}

\end{document}